\documentclass[12pt]{article}
\pdfoutput=1
\usepackage{amssymb,enumerate,amsmath}
\usepackage{empheq}
\usepackage{graphicx}
\usepackage{epsfig}
\usepackage{epstopdf}
\usepackage{latexsym}
\usepackage{psfrag}
\usepackage{booktabs}
\usepackage{bbm}
\usepackage{xspace}
\usepackage[small]{caption}

\usepackage[toc]{appendix}

\usepackage{color}
\usepackage{datetime}
\usepackage[
      colorlinks=false,
      linkcolor=darkblue,  
      urlcolor=blue,    
      filecolor=blue,     
      citecolor=red,
linktocpage=true,
      pdfstartview=FitV,
      bookmarksopen=true    
      ]{hyperref}

\ifpdf
\fi

\numberwithin{equation}{section}  

\def\cG{{\cal G}}

\def\cM{{\cal M}}

\def\cO{{\cal O}}
\def\cP{{\cal P}}
\def\cQ{{\cal Q}}

\def\Neql#1{{\cal N}\!=\!{#1}}

\def\RR{\mathbb{R}}

\definecolor{cardinal}{rgb}{0.6,0,0}
\definecolor{darkgreen}{rgb}{0,0.5,0}
\definecolor{golden}{rgb}{0.92, 0.7, 0}
\definecolor{midnight}{rgb}{0, 0, 0.5}
\definecolor{darkblue}{rgb}{0.2, 0, 0.8}
\definecolor{white}{rgb}{1,1,1}
\definecolor{black}{rgb}{0,0,0}
\definecolor{purple}{rgb}{0.4,0,0.5}

\def\Re{{\rm Re}} \def\Im{{\rm Im}}

\newcommand{\dd}{\mathrm{d}}					
\DeclareMathOperator*{\hodge}{\star}				
\newcommand{\vol}{\mathrm{vol}}				
\DeclareMathOperator{\Lie}{\mathcal{L}}
\DeclareMathOperator{\Ric}{\mathrm{Ric}}			
\DeclareMathOperator{\into}{\raisebox{0.2ex}{\reflectbox{\rotatebox[origin=c]{180}{$\neg$}}}}
\newcommand{\abs}[1]{{\lvert {#1} \rvert}}			

\usepackage{forloop}
\newcounter{ct}

\newcommand{\eucl}[1]{(%
	\ifthenelse{#1 > 0}{%
		\mathord{+} \forloop[-1]{ct}{#1}{\value{ct} > 1}{\, \mathord{+}}%
	}{}%
)}

\newcommand{\sig}[2]{(%
	\ifthenelse{#2 > 0}{%
		\mathord{-}  \forloop[-1]{ct}{#2}{\value{ct} > 1}{\, \mathord{-}}%
		\ifthenelse{#1 > 0}{\,}{}%
	}{}%
	\ifthenelse{#1 > 0}{%
		\mathord{+} \forloop[-1]{ct}{#1}{\value{ct} > 1}{\, \mathord{+}}%
	}{}%
)}

\newcommand{\sref}[1]{Section~\ref{#1}}

\usepackage{empheq}
\newcommand*\padbox[1]{\fbox{\hspace{0.3em}#1\hspace{0.3em}}}
\empheqset{box=\padbox}

\allowdisplaybreaks[1] 

\begin{document}  

\begin{titlepage}
 
\bigskip
\bigskip
\bigskip
\bigskip
\begin{center} 
{\Large \bf  Towards a violation of cosmic censorship}

\medskip
\bigskip
\bigskip
{\bf Benjamin E.~Niehoff, Jorge E.~Santos, Benson Way} \\
\bigskip
DAMTP, Centre for Mathematical Sciences, \\
University of Cambridge\\
Wilberforce Road, Cambridge CB3 0WA, UK\\
\bigskip
\bigskip
{\rm B.E.Niehoff@damtp.cam.ac.uk, ~J.E.Santos@damtp.cam.ac.uk, ~B.Way@damtp.cam.ac.uk} \\

\bigskip
\bigskip 

\end{center}

\begin{abstract}

\noindent Numerical evidence suggests that the superradiant instability of Kerr-AdS black holes and ``black resonators'' evolves to smaller and smaller scales towards a ``limiting" black resonator.  Using $AdS_4$ supergravity, we argue that this putative endpoint to the superradiant instability does not exist.  

\end{abstract}

\end{titlepage}


\tableofcontents

\section{Introduction}
Though AdS was once regarded as an academic curiosity, it has since attracted much more attention due to gauge/gravity duality, where AdS solutions are dual to states of certain large $N$ strongly coupled field theories.  Due to the existence of a timelike boundary, the gravitational dynamics of Anti-de Sitter (AdS) are rather different than that of de Sitter or Minkowski space.  If one imposes energy and momentum conserving boundary conditions, waves can reach infinity in finite time, reflect off the timelike boundary, and return to interact again with the interior. These returning waves can cause black hole formation \cite{DafermosHolzegel2006,Bizon:2011gg,Jalmuzna:2011qw,Buchel:2012uh} or, in the presence of ergoregions, trigger superradiant instabilities \cite{Hawking:1999dp,Kunduri:2006qa,Cardoso:2004hs,Dias:2013sdc,Cardoso:2013pza}.

In an effort to understand the superradiant instability, we investigate the existence of a special class of solutions to the four-dimensional vacuum Einstein equation with negative cosmological constant,
\begin{equation}
R_{ab} = -\frac{3}{L^2}g_{ab}\,,
\label{eq:einstein}
\end{equation}
where $L$ the AdS length scale.  We focus on solutions that are asymptotic to global AdS, where the boundary metric is conformal to the Einstein static universe $\mathbb{R}_t\times S^2$.  We will be preoccupied by stationary\footnote{We call a solution \emph{stationary} if it contains a Killing vector field that is timeline on an open set of the conformal boundary.} solutions, but are ultimately interested in dynamical situations with reflecting boundary conditions, where the energy $E$ and angular momentum $J$ are conserved.  Unlike the Poincar\' e patch, dynamics with this boundary metric and reflecting boundary conditions are non-dissipative.  

Amongst the most relevant stationary solutions to (\ref{eq:einstein}) that satisfy these boundary conditions are of course AdS itself and black holes. The Carter solution \cite{Carter:1968ks}, otherwise known as Kerr-AdS, is the most general black hole with global AdS asymptotics that is known in closed form. Kerr-AdS with angular velocity $\Omega_H L>1$ suffers from the superradiant instability \cite{Hawking:1999dp,Kunduri:2006qa,Cardoso:2004hs,Dias:2013sdc,Cardoso:2013pza}.  This instability has its origins in the Penrose process \cite{Penrose:1971uk}, where energy can be extracted from an ergoregion by particles. The wave analog of this process, where scattered waves can have a larger amplitude than the incident wave, is called superradiance \cite{Zeldovich:1971,Starobinsky:1973,Teukolsky:1974yv}.  For asymptotically flat black holes, superradiant gravitational waves will simply disperse at null infinity.  In AdS, however, the scattered wave returns towards the black hole and extracts more energy.  The process repeats until the (now) high-energy wave backreacts on the geometry, leading to the superradiant instability \cite{Hawking:1999dp,Kunduri:2006qa,Cardoso:2004hs,Dias:2013sdc,Cardoso:2013pza}.  The endpoint of this instability is an open problem in general relativity, which we will partially address.

Recently, a new family of stationary solutions, \emph{black resonators} were found \cite{Dias:2015rxy}.  These are black holes that only have a single (helical) Killing field and can share the same energy and angular momentum as Kerr-AdS.  Though these black holes are entropically favoured over Kerr-AdS, they are still unstable to superradiance, and therefore cannot be the endpoint of this instability.  More precisely, the (yet unpublished) results of \cite{Wald:unpublished} imply that any candidate endpoint must satisfy $\Omega_H L\leq 1$, and all black resonators found in \cite{Dias:2015rxy} have $\Omega_H L>1$.

Actually, for a given energy and angular momentum, there is a countably infinite set of black resonators, a subset of which can be labeled by an integer $m$, with increasing values of $m$ labelling black resonators with increasing (but still bounded) entropy.  Furthermore, (at least for small energy solutions) as $m\rightarrow\infty$, these solutions have an angular frequency that approaches $\Omega_HL\rightarrow1$.  It would be desirable to know if such a ``limiting" $m\to\infty$ black resonator with $\Omega_H L = 1$ exists since it would be a natural candidate for the endpoint of the superradiant instability.

To address this question, we first note that the zero-size limit of black resonators do not approach AdS.  Instead, they connect to horizonless solutions of the Einstein equation called \emph{geons}. These were perturbatively constructed in \cite{Dias:2011ss} and numerically constructed in \cite{Horowitz:2014hja}. Like black resonators, they only have a single helical Killing field, and obey a simplified form of the first law, $\dd E = \Omega \, \dd J$, where $\Omega$ is the angular velocity of the geon \cite{Horowitz:2014hja}. In the limit where black resonators become arbitrarily small, $\Omega_H = \Omega$. In this way, small black resonators can be regarded as a small Kerr-AdS black hole placed in a geon with the same angular frequency.  Such an interpretation leads to quantitative statements \cite{Dias:2011ss,Cardoso:2013pza} (following the methods used in \cite{Basu:2010uz,Bhattacharyya:2010yg,Dias:2011tj,Dias:2011at}) that were tested and confirmed in \cite{Dias:2015rxy}.

It therefore seems natural to suspect that a limiting $\Omega_HL= 1$ black resonator would also be connected to a geon with $\Omega L=1$.  So now we ask whether such a ``limiting" geon exists. From the first law, such a solution satisfies $E = J/L$, which are minimum energy solutions at a given angular momentum $J$. Said another way, these solutions saturate the bound $E \geq J/L$ provided by the positive energy theorem in AdS \cite{Gibbons:1983aq}. One can thus formulate the question in a different manner: are there horizonless solutions of the Einstein equation that are minimum energy, besides AdS itself?

We answer this question in the negative by appealing to supersymmetry.  Although our original theory \eqref{eq:einstein} is not a supersymmetric theory, it still admits \emph{solutions} which are supersymmetric.  An example is empty AdS, which admits four Killing spinors \cite{Gibbons:1983aq}.  More generally, in AdS gravity the positive energy theorem \cite{Gibbons:1983aq} also states that the bound $E \geq J/L$ is saturated if and only if the solution is supersymmetric.  We prove that any supersymmetric solution of \eqref{eq:einstein} which is asymptotically AdS must be AdS itself.%
\footnote{We note that \cite{Gibbons:1983aq} contains a footnote referencing unpublished work \cite{Gibbons:unpub} wherein the same is proven.  In this paper we collect facts from published works and make this proof explicit.}
Thus we argue that the candidate endpoint to the superradiant instability of Kerr-AdS and black resonators does not have a regular zero-size limit.  

Instead, if such a zero-size limit were singular, there must be a singularity located at the centre of a ``limiting" geon that can be covered by a horizon.  We will argue that geons do not develop curvature singularities in their centre in this $m\to\infty$ limit.  The accumulated evidence therefore suggests that the putative endpoint to the superradiant instability does not exist.

This paper is structured as follows.  In \sref{sec:review}, we review black resonators and geons.   In \sref{sec:positive energy}, we then review the arguments of the positive energy theorem in AdS put forth in \cite{Gibbons:1983aq}.  In \sref{sec:4d sugra solns}, we seek putative supersymmetric geons within $\Neql2$ gauged supergravity, by applying the results of \cite{Caldarelli:2003pb,Cacciatori:2004rt} which classify all supersymmetric solutions to this theory.  We show that under the simple assumption of compatibility with pure AdS gravity \eqref{eq:einstein}, the conditions of \cite{Caldarelli:2003pb} lead to a unique local metric tensor.  Imposing asymptotic boundary conditions completes the proof that there are no such geons.  Finally, we discuss the implications of this result.

\section{Geons and Black Resonators}
\label{sec:review}
In this section, we review the phase diagram of black holes in AdS.  Let us begin with a brief discussion on thermodynamic ensembles.  There are typically two ensembles where AdS/CFT is discussed: the microcanonical ensemble and the grand-canonical ensemble. In the microcanonical ensemble, the energy and angular momentum of the system are kept fixed and the relevant thermodynamic potential is the entropy. In the grand-canonical ensemble, the temperature $T$ and angular velocity are kept fixed, and the relevant thermodynamic potential is the Gibbs free energy $G \equiv E-\Omega_H J-T \mathcal{A}/4$, where $\mathcal{A}$ is the area of the spatial cross section of the horizon.

Since we wish to discuss the relevant scenario where energy and angular momentum are conserved, we will focus mainly on the microcanonical ensemble.  In this ensemble, black holes are preferred over pure AdS due to their entropy.  In the context of AdS/CFT, black holes have an entropy that scales as $N^2$ for large $N$, while ``empty'' AdS has an entropy\footnote{This entropy comes from stringy modes \cite{Horowitz:1996nw}.} that scales as $N$ \cite{Horowitz:1996nw}.

Let us now review the construction of geons.  Recall that the spectrum of linear gravitational perturbations of AdS \cite{Dias:2011ss,Dias:2012tq} is labeled by a type (scalar or vector), and several wavenumbers.  These wavenumbers are the polar wavenumber $\ell$, the azimuthal wavenumber $m$, and a radial wavenumber $p$.   The frequency is given by integer normal modes
\begin{equation}
\omega_{AdS}L=s+\ell+2p\;,
\end{equation}
where $s=1$ for scalars and $s=2$ for vectors.  The wavenumbers $\ell$ and $m$ come from spherical harmonics and satisfy the relation $\ell\geq|m|$.  Due to the presence of discrete symmetries, we henceforth take $m$ to be positive without loss of generality.

One can attempt to start with some linear combination of these modes and continue to higher orders in perturbation theory.  Generically, more than one mode is excited, and the fact that these normal modes are evenly spaced causes higher-order modes to be excited that grow linearly in time.  This leads to a breakdown of perturbation theory and is conjectured to precede dynamical black hole formation (a nonlinear instability of AdS) \cite{DafermosHolzegel2006,Bizon:2011gg,Dias:2011ss,Dias:2012tq}.  

However, exciting only a single-mode gives no obstruction to perturbation theory. Perturbation theory survives to arbitrarily high order and yields a perturbative construction of a new family of horizonless time-periodic solutions -- the geons \cite{Dias:2011ss}.  There is therefore a one-parameter family of geons for every normal mode of AdS. At lowest order in perturbation theory, one finds $EL=\omega_{AdS}J/m$. A non-perturbative numerical construction of geons can be found in \cite{Horowitz:2014hja}. These geons have a characteristic angular frequency $\Omega$ and contain only a single Killing field $K=\partial_t+\Omega \partial_\phi$, where $\partial_t$ and $\partial_\phi$ are asymptotically the time and azimuthal coordinates on the boundary.  Geons have finite energy $E$ and angular momentum $J$ and approach AdS in the limit of small $E$ or small $J$. 

Black resonators can be found by performing a matched asymptotic expansion to place a small Kerr-AdS black hole on a geon background \cite{Dias:2011ss,Cardoso:2013pza}. The angular frequency of the small Kerr-AdS black hole must match the frequency of the geon $\Omega_H=\Omega$, to ensure that there is no energy flux across the black hole horizon.  For black resonators with small $E$ and $J$, one can proceed perturbatively and obtain an approximate expression for the entropy \cite{Dias:2011ss,Cardoso:2013pza}:
\begin{equation}\label{eq:entropy}
S=4\pi E^2\left(1-\frac{\omega_{AdS}}{m}\frac{J}{EL}\right)^2\;,
\end{equation}
where the quantity inside the parentheses is guaranteed to be positive by the first law. Let us attempt to maximise this entropy at a fixed $E$, $J$, which is equivalent to minimising $\omega_{AdS}/m$.  Note that we have the bound
\begin{equation}
\omega_{AdS}L/m=\frac{s+\ell+2p}{m}\geq\frac{\ell}{m}\geq 1\;.
\end{equation}
This bound is saturated in the limit $m\rightarrow\infty$, $\ell\rightarrow\infty$, keeping $\ell/m=1$, and $p$ finite, as first noted in \cite{Kunduri:2006qa}. For definitiveness, we will later restrict ourselves to the $s=1$, $p=0$ and $\ell=m$ family, and consider the $m\rightarrow\infty$ limit.  Our conclusions will be the same for any other limit that has $\omega_{AdS}L/m\rightarrow 1$.

Black resonators are also solutions that branch from the onset of the superradiant instability in Kerr-AdS \cite{Dias:2015rxy}. That is, they are black holes that connect the onset of superradiance to the geons  \cite{Dias:2015rxy}.  Let us therefore review how the onset of this instability was found \cite{Dias:2013sdc,Cardoso:2013pza}. One proceeds via linear perturbation theory about the Kerr-AdS metric $\bar g$.  Recall that Kerr-AdS is a two-parameter family that can be uniquely parametrised by its energy $E$ and angular momentum $J$. Alternatively, this solution can also be parametrised by the angular velocity with respect to the boundary $\Omega_H$, and the area radius $R_+ \equiv \sqrt{\mathcal{A}/4\pi}$, where $\mathcal{A}$ is the horizon area.  The Kerr-AdS black hole has two commuting Killing fields, $\partial_t$ and $\partial_\phi$, as well as the so-called $t$-$\phi$ symmetry which is a discrete symmetry that acts according to $(t,\phi)\to-(t,\phi)$.

Now consider linear fluctuations $g_{ab} = \bar{g}_{ab}+e^{-i\omega t+im\phi}h_{ab}$, where we have used the Killing fields $\partial_t$ and $\partial_\phi$ of Kerr-AdS to place the dependence on $t$ and $\phi$ in a Fourier mode decomposition. The linearised Einstein equation becomes a complicated set of PDEs in the form of a quadratic eigenvalue problem in $\omega$.  For boundary conditions, one chooses ingoing conditions at the horizon, and normalisable (finite energy) modes at the boundary.  This system can be solved using the Newman-Penrose formalism, and by writing the natural boundary conditions for the metric $h_{ab}$ in terms of boundary conditions for the so-called Teukolsky scalars \cite{Teukolsky:1973ha}.  

Like the normal modes of AdS, the modes of Kerr-AdS are labeled by a type (scalar or vector), a polar wavenumber $\ell$, an azimuthal wavenumber $m$, and a radial wavenumber $p$; so the frequencies can be parametrised in full as $\omega(s,\ell,m,p,E,J)$, or equivalently $\omega(s,\ell,m,p,\Omega_H,R_+)$. The onset of the superradiant instability occurs precisely when $\Re(\omega) = m\,\Omega_H$.  Therefore, given $s$, $\ell$, $m$, and $p$, there is a continuous one-parameter family of onsets.  The behaviour of these onsets is intricate, with different families crossing each other in phase space \cite{Cardoso:2013pza}.  

Black resonators branch from these onsets and connect (in moduli space) to the geons.  Each $s$,$\ell$,$m$,$p$ therefore yields a two-parameter family of black resonators (which can be parametrised by $E$, and $J$, or $\Omega_H$ and $R_+$).  Black resonators were numerically constructed beyond perturbation theory in \cite{Dias:2015rxy}.

\begin{figure}[t]
\centering
\includegraphics[height = 0.4\textheight]{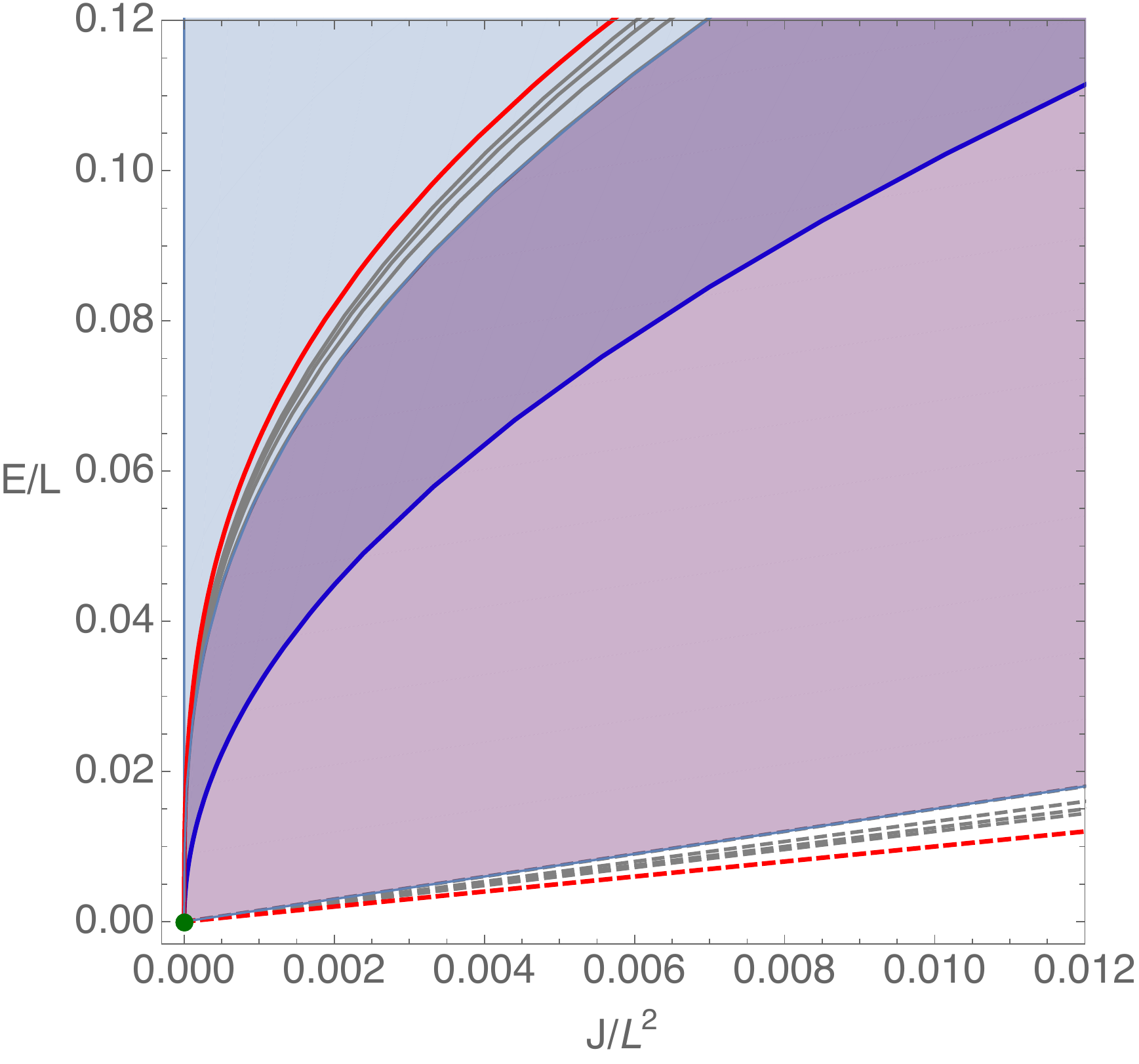}
\caption{\label{fig:1} 
$E$ vs. $J$ phase diagram of $AdS_4$ solutions.  Pure $AdS_4$ is the green dot at the origin. Kerr-AdS lies in the light blue region above the solid blue line, which represents extremal Kerr-AdS. The dotted lines near the bottom are geons ($s=1$, $p=0$, and $\ell=m=2,\ldots,4$ shown), with the thick red line being the ``limiting" geon family.  The thin lines are the onsets of the superradiant instability (only modes $s=1$, $p=0$, and $\ell=m=2,\ldots,4$ shown), with the thick red line being the $\Omega_HL=1$ curve where superradiance shuts down for all modes.  Black resonators with $s=1$, $p=0$, and $\ell=m=2$ lie in the purple region.  There are black resonators for higher modes as well, and they all connect the onsets to the geons.}
\end{figure}

Now let us put these pieces together and consider the full phase diagram in the regime of small energy and angular momentum where we have perturbative control.  As mentioned earlier, let us also restrict ourselves to the $s=1$ (scalar), $p=0$ and $\ell=m$ family of black resonators and geons. This phase diagram can be seen in Figure \ref{fig:1}.   Here, the onset modes, the black resonators, and geons all have an angular frequency approximately given by $\Omega_HL\approx\Omega\approx1+1/m$.  For a given $m$, the $E$ and $J$ values of the onsets are approximately given by the $E$ and $J$ values of the Kerr-AdS black hole with $\Omega_HL=1+1/m$; the geons lie approximately at $E/L=(1+1/m)J/L^2$; and the black resonators lie in the space between these two curves.  

For the lowest value of $m=2$, these black resonators were found nonlinearly \cite{Dias:2015rxy}, and their entropy is higher than that of Kerr-AdS for the same $E$ and $J$.  From \eqref{eq:entropy}, black resonators with the same $E$ and $J$ have increasing entropy with increasing $m$, but the entropy remains bounded.  Furthermore, all of these black resonators have $\Omega_H L>1$.  Therefore, they must be unstable to superradiance.  More precisely, the (yet unpublished) results of \cite{Wald:unpublished} mathematically prove that these black resonators are unstable, since no Killing vector field that is everywhere timelike can be found at the conformal boundary.

But, a number of interesting things happen in the limit $m\rightarrow\infty$.  First, $\Omega_HL\rightarrow1$, which shuts down the superradiant instability.  That is, a ``limiting" black resonator is stable to superradiance and, as we pointed out earlier, is entropically dominant to all other black resonators of the same $E$ and $J$.  If such a solution exists, it is therefore a prime candidate for the endpoint of the superradiant instability.  Second, note also that the ``limiting" geon satisfies $E/L=J/L^2$.  That is, these geons minimise the energy for a given angular momentum and therefore must be a supersymmetric solution (we will discuss the positivity of energy theorem in the next section).  Since black resonators always connect to geons, at least for the asymptotic charges probed in \cite{Dias:2015rxy}, we can address the existence of a limiting black resonator by attempting to address the existence of a limiting geon. 

Our argument that limiting geons saturate the bound $E/L\geq J/L^2$ relies on perturbation theory.  It is possible that this is no longer true when considering the fully nonlinear geons due to small corrections in the perturbation expansion.  However, the bound must still be satisfied (as we will discuss in the following section), so the first law for geons $\dd E = \Omega \, \dd J$ implies that $\Omega\geq1$.  That is, small black holes placed in a geon have $\Omega_H L\geq1$, and so must be superradiant unless the bound $E/L\geq J/L^2$ is saturated.  It is therefore sufficient for our purposes to assume that limiting geons satisfy $E/L=J/L^2$.

\section{The Positive Energy Theorem in AdS}
\label{sec:positive energy}

As we pointed out in the previous section, the $m\rightarrow\infty$ limit of geons satisfies
\begin{equation} \label{endpoint}
E = \abs{J} / L,
\end{equation}
and we are interested in whether such a limiting solution exists.  In \cite{Gibbons:1983aq}, it is proven that generic asymptotically-AdS solutions to gravity theories with negative cosmological constant, whose energy-momentum tensor satisfies the dominant energy condition,%
\footnote{The dominant energy condition states that if $u^\mu$ is either timelike or null, and future-pointing, then $-T^\mu{}_\nu u^\nu$ is also either timelike or null, and future-pointing.}
are subject to the bound
\begin{equation} \label{BPS bound}
E \geq \abs{J}/L,
\end{equation}
and that furthermore, this bound is saturated if and only if the solution admits a \emph{Killing spinor}.  Therefore the limiting geon we seek, if it exists, must be a \emph{supersymmetric} solution to AdS gravity.  In this section, we review the proof of the bound \eqref{BPS bound} and its relation to supersymmetry.

The authors of \cite{Gibbons:1983aq} employ a similar line of reasoning to \cite{Witten:1981mf}, which makes use of various properties of Killing spinors.  A Killing spinor is a (non-trivial) spinor $\epsilon$ which is covariantly constant on the spacetime manifold $\cM$,
\begin{equation} \label{Killing spinor eqn}
\hat \nabla_\mu \epsilon = 0,
\end{equation}
for some notion of covariant derivative (which includes the Levi-Civit\`a connection in the spin representation, as well as possibly terms coming from gauge connections and the cosmological constant).  Given a solution to \eqref{Killing spinor eqn}, one can construct a variety of quantities out of spinor bilinears,\footnote{Here $\bar \epsilon$ is some appropriately-defined conjugate spinor.}
\begin{equation}
V^\mu \equiv \bar \epsilon \gamma^\mu \epsilon, \qquad \Phi^{\mu\nu} \equiv \bar \epsilon \gamma^{[\mu} \gamma^{\nu]} \epsilon, \qquad \text{etc.},
\end{equation}
which then inherit from \eqref{Killing spinor eqn} a number of first-order differential relations.  In particular, the vector $V^\mu$, so constructed, is always a Killing vector, and justifies our calling spinor solutions to \eqref{Killing spinor eqn} ``Killing''.  Using Fierz identities, one can also show that a number of algebraic relations hold, including (taking the metric signature to be ``mostly plus'')
\begin{equation} \label{V cond}
V^\mu V_\mu \leq 0,
\end{equation}
and thus $V^\mu$ is timelike or null.

To obtain the energy bound, we will find it useful to consider AdS as a 4-dimensional submanifold of $\RR^{3,2}$ defined by the quadric surface
\begin{equation} \label{quadric}
(X^0)^2 - (X^1)^2 - (X^2)^2 - (X^3)^2 + (X^4)^2 = - \frac{3}{\Lambda},
\end{equation}
where $\Lambda = -6/L^2$ is the cosmological constant.  We then consider, as discussed in \cite{Abbott:1981ff}, a collection of currents $J_{AB}$ which generate the $O(3,2)$ group under which \eqref{quadric} is invariant.  Then $J_{04}$ corresponds to the energy $E$, and $J_{13}$ corresponds to the angular momentum $J$.  The remaining $J_{AB}$ can be made to vanish by an appropriate $O(3,2)$ rotation.

Choosing $\epsilon$ to be a 4-component commuting Dirac spinor, the appropriate covariant derivative in the presence of a (negative) cosmological constant is
\begin{equation} \label{nabla def}
\hat \nabla_\mu \epsilon \equiv \nabla_\mu \epsilon + i \big( - \tfrac{1}{12} \Lambda \big)^{1/2} \gamma_\mu \epsilon,
\end{equation}
where we shall match the gamma-matrix conventions of \cite{Gibbons:1983aq}.  With respect to this covariant derivative, it is easy to show that empty AdS spacetime admits four linearly-independent Killing spinors.  We are then interested in more general spacetimes which are asymptotically AdS.  While these spacetimes will not admit Killing spinors in general, they must admit \emph{asymptotic} solutions to \eqref{Killing spinor eqn}, in the sense that
\begin{equation} \label{asym cond}
\hat \nabla_\mu \epsilon = \cO(1/r^2), \qquad \text{as} \; r \to \infty,
\end{equation}
as this is merely the condition that a solution approach AdS sufficiently fast.  The spinor $\epsilon$ itself approaches
\begin{equation} \label{epsilon 0}
\epsilon = s \epsilon_0 + \cO(1/r),
\end{equation}
where $\epsilon_0$ is a constant, non-zero spinor and $s$ satisfies $\bar s s = \mathbf{1}$.

Following \cite{Nester:1982tr}, one can construct out of $\epsilon$ an antisymmetric tensor
\begin{equation} \label{integrand}
E^{\mu\nu} \equiv 2 \big( \bar \epsilon \gamma^{\mu\nu\rho} \hat \nabla_\rho \epsilon - \overline{\hat \nabla_\rho \epsilon} \gamma^{\mu\nu\rho} \epsilon \big), \qquad \gamma^{\mu\nu\rho} \equiv \gamma^{[\mu} \gamma^\nu \gamma^{\rho]},
\end{equation}
such that the $J_{AB}$ are given by the surface integral of $E_{\mu\nu}$ on the $S^2$ at infinity:
\begin{equation}
2 i \bar \epsilon_0 J_{AB} \sigma^{AB} \epsilon_0 = \frac12 \oint_{S^2_\infty} E_{\mu\nu} \, \dd \Sigma^{\mu\nu},
\end{equation}
where $\sigma^{AB}$ are the generators of $SO(3,2)$ in the spin representation.  One can then apply Stokes' theorem to obtain
\begin{equation} \label{stokes}
2 i \bar \epsilon_0 J_{AB} \sigma^{AB} \epsilon_0 = \int_{\Sigma} \nabla_\nu E^{\mu\nu} \, \dd \Sigma_\mu,
\end{equation}
where now the integral is over the entire hyperslice $\Sigma$.  Surface terms (such as at internal horizons) can be avoided by taking $\Sigma$ to be a smooth 3-surface which is everywhere spacelike.  The divergence of \eqref{integrand} is
\begin{equation}
\nabla_\nu E^{\mu\nu} = 2 \Big[ \overline{\hat \nabla_\nu \epsilon} \gamma^{\mu\nu\rho} \hat \nabla_\rho \epsilon + \bar \epsilon \gamma^{\mu\nu\rho} \hat \nabla_\nu \hat \nabla_\rho \epsilon \Big] + \mathrm{c.c.},
\end{equation}
and the second term, after using the Ricci identity and the Einstein equation, can be written in terms of the energy-momentum tensor of all the matter fields:
\begin{equation}
\bar \epsilon \gamma^{\mu\nu\rho} \hat \nabla_\nu \hat \nabla_\rho \epsilon = \frac12 T^{\mu\nu}_{\text{mat}} \, \bar \epsilon \gamma_\nu \epsilon.
\end{equation}
Thus the integral \eqref{stokes} becomes
\begin{equation} \label{energy integral}
\frac12 i \bar \epsilon_0 J_{AB} \sigma^{AB} \epsilon_0 = \int_{\Sigma} \Big[ \overline{\hat \nabla_\nu \epsilon} \gamma^{\mu\nu\rho} \hat \nabla_\rho \epsilon + \frac12 T^{\mu\nu}_{\text{mat}} \, \bar \epsilon \gamma_\nu \epsilon \Big] \, \dd \Sigma_\mu.
\end{equation}
As stated earlier, the vector $V_\mu \equiv \bar \epsilon \gamma_\mu \epsilon$ is always timelike or null.  Therefore if $T^{\mu\nu}_\text{mat}$ satisfies the dominant energy condition, then the second term in \eqref{energy integral} is always non-negative.  This establishes that $J_{AB}$ is bounded by
\begin{equation} \label{4d bound}
\frac12 i \bar \epsilon_0 J_{AB} \sigma^{AB} \epsilon_0 \geq \int_{\Sigma} \Big[ \overline{\hat \nabla_\nu \epsilon} \gamma^{\mu\nu\rho} \hat \nabla_\rho \epsilon \Big] \, \dd \Sigma_\mu.
\end{equation}

To complete the proof, one chooses orthonormal frames such that the 0 direction is normal to $\Sigma$.  In this basis, the integral \eqref{4d bound} can be expanded as (indices $a,b$ are 3-dimensional indices along $\Sigma$):
\begin{equation}
\frac12 i \bar \epsilon_0 J_{AB} \sigma^{AB} \epsilon_0 \geq 2 \int_{\Sigma} \Big[ (\hat \nabla^a \epsilon)^\dag (\hat \nabla_a \epsilon) - (\gamma^a \hat \nabla_a \epsilon)^\dag (\gamma^b \hat \nabla_b \epsilon) \Big] \, \dd \Sigma_0.
\end{equation}
One then observes that on any smooth spacelike surface $\Sigma$, the ``Witten condition"
\begin{equation}
\gamma^a \hat \nabla_a \epsilon = 0
\end{equation}
is an elliptic equation and has no zero modes.  Thus there always exists a solution on $\Sigma$ satisfying the boundary conditions \eqref{asym cond}, \eqref{epsilon 0} for a given $\epsilon_0$.  Choosing this $\epsilon$, we then have
\begin{equation}
\frac12 i \bar \epsilon_0 J_{AB} \sigma^{AB} \epsilon_0 \geq 2 \int_{\Sigma} (\hat \nabla^a \epsilon)^\dag (\hat \nabla_a \epsilon) \, \dd \Sigma_0,
\end{equation}
and, since $\epsilon_0$ was arbitrary, the matrix (in the spinor indices) sandwiched between $\epsilon_0^\dag (\ldots) \epsilon_0$ on the left-hand side is seen to be positive semidefinite.  Taking its trace, we obtain
\begin{equation}
\abs{J_{04}} - \abs{J_{13}} \geq 0,
\end{equation}
and thus
\begin{equation} \label{final bound}
E \geq \abs{J} /L.
\end{equation}
Furthermore, we see that \eqref{final bound} is saturated only if\footnote{After proving that the unique solution with Killing spinors is AdS, we can strengthen this to ``if and only if''.}
\begin{equation}
\hat \nabla_a \epsilon = 0.
\end{equation}
This is an equation only on $\Sigma$; however, by choosing a slightly different $\Sigma$ with the same boundary, one can show that $\epsilon$ must in fact solve the full Killing spinor equation on $\cM$:
\begin{equation} \label{Killing spinor 2}
\hat \nabla_\mu \epsilon = 0.
\end{equation}
Thus one has shown that, in asymptotically-AdS spacetimes, the energy is bounded from below by the angular momentum \eqref{final bound}, and this bound is saturated only if the spacetime admits a Killing spinor.

In \cite{Gibbons:1983aq} it is furthermore shown that if $E = \abs{J}/L = 0$, then one in fact has \emph{four} Killing spinors, and the spacetime must be AdS itself.  If $E = \abs{J}/L = 0$, then we must have that $i \bar \epsilon_0 J_{AB} \sigma^{AB} \epsilon_0 = 0$ for \emph{all} choices of $\epsilon_0$.  This implies that we have a complete basis of Killing spinors.  From \eqref{Killing spinor 2} one can obtain an integrability condition for Killing spinors to exist:
\begin{equation}
\hat \nabla_{[\mu} \hat \nabla_{\nu]} \epsilon = \frac14 \big[ R_{\mu\nu\rho\sigma} + \frac23 \Lambda \, g_{\mu\rho} g_{\nu\sigma} \big] \sigma^{\rho\sigma} \epsilon = 0.
\end{equation}
And if \eqref{Killing spinor 2} admits a complete basis of Killing spinors, we must have
\begin{equation}
\big[ R_{\mu\nu\rho\sigma} + \frac23 \Lambda \, g_{\mu\rho} g_{\nu\sigma} \big] \sigma^{\rho\sigma} = 0,
\end{equation}
which implies that the spacetime has constant curvature.  The asymptotic boundary conditions then imply that the spacetime is AdS.

However, this argument does not apply to our situation, where we assume only that $E = \abs{J}/L$, which \emph{a priori} can be saturated for any $J$.  Indeed, geons always have $J\neq0$.  In the following section, we develop a more general argument that relies on weaker assumptions.

\section{Supersymmetric Solutions of 4D Gauged Supergravity}
\label{sec:4d sugra solns}

To prove that the only asymptotically-AdS solution to \eqref{eq:einstein} with Killing spinors is AdS, we approach the problem from $\Neql2$ gauged supergravity.  To a purist studying the problem \eqref{eq:einstein}, this may seem like an extraneous maneuver; supergravity theories contain numerous additional fields, such as gauge fields and scalars, and their fermionic superpartners.  But a supergravity theory can inform us about supersymmetric solutions to pure Einstein gravity by choosing these additional fields to be trivial.%
\footnote{Trivial in this case means setting the scalars of the gravity multiplet to constants and other fields to zero.}

We choose $\Neql2$ gauged supergravity because a simple classification exists in the literature \cite{Cacciatori:2007vn} of its supersymmetric classical configurations.  The field content is solely that of the $\Neql2$ gravity multiplet, consisting of a graviton, two gravitini, and a Maxwell gauge field:
\begin{equation}
g_{\mu\nu}, \quad \psi^i_\mu, \quad A_\mu, \qquad i  = 1,2,
\end{equation}
and the action is that of Einstein-Maxwell-$\Lambda$ plus Fermi terms.  We are interested in classical solutions (thus $\psi^i_\mu = 0$) where the gauge field $A_\mu = 0$, and thus these solutions also solve \eqref{eq:einstein}.

In \cite{Cacciatori:2007vn} the authors work out the conditions which must be satisfied in order for a classical solution be supersymmetric.  These are essentially the existence of a Killing spinor
\begin{equation} \label{Killing spinor 3}
\hat \nabla_\mu \epsilon = 0,
\end{equation}
where $\hat \nabla_\mu$ contains the usual cosmological term as in \eqref{nabla def}, as well as a twist constructed out of the `supercovariant' gauge field strength $\hat F_{\mu\nu} \equiv \partial_\mu A_\nu - \partial_\nu A_\mu - \Im (\bar \psi_\mu \psi_\nu)$.  To obtain a useful set of supersymmetry conditions, one constructs various spinor bilinears out of $\epsilon$ and explores the consequences of \eqref{Killing spinor 3} and the Fierz identities.

Here we summarize the results of \cite{Caldarelli:2003pb} without delving into too much detail.  Out of spinor bilinears, one can construct a scalar $\varphi$, a pseudoscalar $\varrho$,%
\footnote{Here we have re-named the functions $f$ and $g$ of \cite{Caldarelli:2003pb} to avoid confusion with the metric, which we call $g$, and a conformal rescaling, which we shall call $f$.}
a vector $V$, a pseudovector $A$, and a 2-form $\Phi$.  Together with the Maxwell 2-form $F$, these satisfy the algebraic relations, obtained from Fierz identities
\begin{align}
\varphi \, V &= - \hodge_4 \, (A \wedge \Phi), & \varphi \, A &= - \hodge_4 \, (V \wedge \Phi), \\
\varrho \, V &= - (\vec A \into \Phi), & \varrho \, A &= - (\vec V \into \Phi), \\
\vec A \cdot \vec V &= 0, & A^2 &= -V^2 = \varphi^2 + \varrho^2 \equiv -N, \\
\varphi \varrho &= - \frac12 \hodge_4 \, (\Phi \wedge \Phi), & \varphi \, \Phi - \varrho \hodge_4 \Phi &= - \hodge_4 \, (V \wedge A),
\end{align}
\begin{equation}
\Phi_{(\mu}{}^\rho \varepsilon_{\nu)\rho\alpha\beta} \Phi^{\alpha\beta} - \frac14 g_{\mu\nu} \varepsilon_{\rho\sigma\alpha\beta} \Phi^{\rho\sigma} \Phi^{\alpha\beta} = 0.
\end{equation}
One also has differential relations obtained from \eqref{Killing spinor 3}:
\begin{gather}
\dd \varphi = - (\vec V \into F), \qquad \dd \varrho + \frac1L A = - \hodge_4 \, (V \wedge F), \\
\dd A = 0, \qquad \dd V = 2 \, \Big( \frac1L \Phi - \varphi \, F - \varrho \hodge_4 F \Big), \\
\Lie_{\vec V} g = 0, \qquad \Lie_{\vec V} F = 0, \qquad \Lie_{\vec A} g = - \frac2L \, \varrho \, g - 2 \hat T, \\
\nabla_\mu \Phi_{\alpha\beta} = \frac{2}{L} g_{\mu[\alpha} V_{\beta]} + 2 F_{[\alpha}{}^\sigma \varepsilon_{\beta] \sigma\mu\nu} A^\nu + F_\mu{}^\sigma \varepsilon_{\sigma\alpha\beta\nu} A^\nu + g_{\mu [\alpha} \varepsilon_{\beta] \nu\rho\sigma} A^\nu F^{\rho\sigma},
\end{gather}
where $\hat T$ is an energy-momentum-like tensor given by
\begin{equation}
\hat T_{\mu\nu} \equiv F_{(\mu}{}^\rho \varepsilon_{\nu)\rho\alpha\beta} \Phi^{\alpha\beta} - \frac14 g_{\mu\nu} \varepsilon_{\rho\sigma\alpha\beta} F^{\rho\sigma} \Phi^{\alpha\beta}.
\end{equation}

As seen in \eqref{V cond}, the Killing vector $V$ may be either timelike or null.  One can attempt to argue whether we must make one choice or the other.  The geon solutions of \cite{Horowitz:2014hja} have wedge-shaped regions, distributed in a ``fan'' around the central axis, where their unique Killing vector is alternately timelike and spacelike.  One could argue that since finite regions of timelike Killing vector exist, we must choose $V$ timelike.  On the other hand, it is unclear what should happen in the $m \to +\infty$ limit.  This limit induces highly oscillatory behaviour as one circles the axis, and one could argue that the timelike and spacelike regions ``average out'', leaving a null Killing vector.%
\footnote{Except on the axis itself, where the Killing vector is always timelike, but this is a set of measure zero.}
As such, we see no convincing argument that the limiting geon, if it exists, must be in one class or the other.  For completeness, we examine both the timelike and null cases:

\subsection{The Timelike Case}

First, take $V$ to be timelike, and hence $N < 0$.  The authors of \cite{Caldarelli:2003pb} provide a construction of timelike solutions; however, to clarify our particular case (where the Maxwell field $F = 0$), we find it more straightforward to provide our own construction.

We can choose $\vec V = \partial_t$, and write the metric ansatz as
\begin{equation}
g = N (\dd t + \omega)^2 - \frac{1}{N} \, \dd s_3^2,
\end{equation}
where everything is independent of $t$.%
\footnote{Here we will make a departure from \cite{Caldarelli:2003pb} and not set any pre-emptive ansatz for $\dd s_3^2$, aside from that it be independent of $t$.}
Then the above collection of supersymmetry conditions imply that $F$ and $\Phi$ can be written:
\begin{align}
\Phi &= - \frac{1}{N} \bigg[ \varrho \, V \wedge A - \varphi \hodge_4 \, (V \wedge A) \bigg], \\
F &= - \frac{1}{N} \bigg[ V \wedge \dd \varphi + \hodge_4 \, \Big( V \wedge \big( \dd \varrho + \frac1L \, A \big) \Big) \bigg],
\end{align}
where as a 1-form, $V = \dd t + \omega$.

\subsubsection{Purely gravitational solutions}

We are interested in purely gravitational solutions, where $F = 0$.  Therefore the various differential conditions simplify to
\begin{equation} \label{grav diff rel}
\dd \varphi = 0, \qquad \dd \varrho + \frac1L \, A = 0, \qquad \dd A = 0, \qquad \dd V = \frac2L \, \Phi, \qquad \Lie_{\vec A} g = - \frac2L \, \varrho \, g.
\end{equation}
The last of these tells us that $\vec A$ is a \emph{conformal} Killing vector.  Therefore it is possible to define a conformally rescaled metric
\begin{equation}
\hat g = e^{2f} \, g
\end{equation}
on which $\vec A$ is an ordinary Killing vector.  Moreover, since the function $\varrho$ is independent of $V$, then we conclude that $V$ is \emph{also} a Killing vector of $\hat g$, and $A$ and $V$ commute.  Thus we can reduce the problem to cohomogeneity 2.

First we find the appropriate conformal rescaling $f$.  Choose $\vec A = \partial_z$ for some coordinate $z$ (not necessarily the same one as in \cite{Caldarelli:2003pb}).  We want
\begin{equation}
\Lie_{\vec A} \hat g = 0 = 2 e^{2f} f_z \, g - e^{2f} \Big( \frac2L \, \varrho \, g \Big),
\end{equation}
and hence we must choose
\begin{equation} \label{fz mu}
f_z = \frac1L \varrho.
\end{equation}
Now we write down a metric ansatz for $\hat g$.  Since $V$ and $A$ commute and $\vec V \cdot \vec A = 0$ (which is preserved under conformal rescaling), we can write
\begin{equation}
\hat g = - \cP \, (\dd t + \omega)^2 + \cQ \, (\dd z + \beta)^2 + h,
\end{equation}
where $h$ is a 2-dimensional metric in some coordinates $(x,y)$, and all quantities depend only on $(x,y)$.  Re-inserting the conformal factor, we have $g = e^{-2f} \, \hat g$, and thus
\begin{equation}
g = e^{-2f} \bigg[ - \cP \, (\dd t + \omega)^2 + \cQ \, (\dd z + \beta)^2 + e^{2u} (\dd x^2 + \dd y^2) \bigg],
\end{equation}
where now $f$ may depend on $z$, but all other quantities are functions of $x, y$ only.

From the algebraic supersymmetry conditions, we conclude that $\cP = \cQ$.  Now let us implement the differential conditions.  First $\dd A = 0$ gives
\begin{equation}
-2 \cQ \, (f_z \, \dd z + \hat \dd f) \wedge (\dd z + \beta) + \hat \dd \cQ \wedge (\dd z + \beta) + \cQ \, \hat \dd \beta = 0,
\end{equation}
where we define $\hat \dd$ to act only on $x, y$.  Hence it must be true that $\hat \dd \beta = 0$, and thus $\beta$ can be eliminated by a shift in the definition of $z$.  What remains is
\begin{equation}
\hat \dd (\log \cQ) \wedge \dd z = 2 \, \hat \dd f \wedge \dd z.
\end{equation}
But the right-hand side generically depends on $z$, while the left-hand side does not.  Thus we are left with two possibilities:
\begin{enumerate}[i.]
\item \emph{Case 1}:  In fact $f_z = 0$, and thus $2f = \log \cQ$, or
\item \emph{Case 2}:  $f_z \neq 0$, and thus $\hat \dd f = \hat \dd \cQ = 0$.  That is, $\cQ$ is a constant, and $f$ is a function of $z$ only.
\end{enumerate}
In the first case, by \eqref{fz mu} one would have $\varrho = 0$, and then by the second relation in \eqref{grav diff rel}, one would have $A = 0$.  This is clearly inconsistent with supersymmetry, which requires $A$ to be non-vanishing.  Therefore the first case is eliminated.

Proceeding with the second case, then, we can choose $\cP = \cQ = 1$ without loss of generality, and $f = f(z)$ only.  The metric ansatz becomes
\begin{equation}
g = e^{-2f} \bigg[ - (\dd t + \omega)^2 + \dd z^2 + e^{2u} (\dd x^2 + \dd y^2) \bigg].
\end{equation}

\subsubsection{Choosing Another Gauge}

To proceed further, it is useful to choose a different gauge.  The metric contains a term $e^{-2f(z)} \, dz^2$, which we may easily change into any form we like.  First we observe that
\begin{equation}
\dd \varphi = 0, \qquad \text{hence} \qquad \varphi \equiv \varphi_0 = \text{const}.
\end{equation}
Next we use $A^2 = \varphi^2 + \varrho^2$, which gives
\begin{equation}
e^{-2f} = \varphi_0^2 + \varrho^2.
\end{equation}
Keeping this in mind, we can also approach from the other direction.  Combining the facts
\begin{equation}
\dd \varrho + \frac1L \, A = 0, \qquad \text{and} \qquad f_z = \frac1L \, \varrho,
\end{equation}
we can integrate the first equation and obtain again $e^{-2f} = \varrho^2 + c$.  Thus we set $c = \varphi_0^2$, and we can write the metric as
\begin{equation} \label{simple g}
g = \frac{L^2 \, d\varrho^2}{\varrho^2 + \varphi_0^2} + (\varrho^2 + \varphi_0^2) \, \bigg[ - (\dd t + \omega)^2 + e^{2u(x,y)} \, (\dd x^2 + \dd y^2) \bigg].
\end{equation}
It is also useful to write down that, in this gauge, we have
\begin{equation}
-N = \varrho^2 + \varphi_0^2, \qquad A = - L \, \dd \varrho, \qquad V = - (\varrho^2 + \varphi_0^2) \, (\dd t + \omega).
\end{equation}

There is one remaining supersymmetry condition to impose, given by
\begin{equation}
\dd V = \frac2L \, \Phi = - \frac2L \frac{1}{N} \bigg[ \varrho \, V \wedge A - \varphi \hodge_4 \, (V \wedge A) \bigg]
\end{equation}
Taking the exterior derivative of this results in the condition
\begin{equation} \label{d omega}
\dd \omega = - \frac2L \, \vol_h = - \frac2L \, e^{2u} \, \dd x \wedge \dd y.
\end{equation}
We point out that on a 2-surface with metric $h$, one always has a complex structure and $\vol_h$ is its K\"ahler 2-form.  Thus the angular momentum $\omega$ is $-2/L$ times the 1-form potential of the K\"ahler form.

\subsubsection{Solving the Einstein equation}

The supersymmetry conditions we have solved thus far are not sufficient to solve the equations of motion.  We must also use the Einstein equation
\begin{equation}
\Ric_g - \frac12 R_g \, g + \Lambda \, g = 0.
\end{equation}
We will show that this is sufficient to fix the constant $\varphi_0$ and the 2-metric
\begin{equation}
h \equiv e^{2u} \, ( \dd x^2 + \dd y^2).
\end{equation}
To begin, we find it useful to define
\begin{equation}
\varrho = \varphi_0 \, \sinh \rho.
\end{equation}
Then the metric ansatz \eqref{simple g} becomes
\begin{equation} \label{g rho}
g = L^2 \, \dd \rho^2 + \varphi_0^2 \cosh^2 \rho \, \Big( - (\dd t + \omega)^2 + h \Big).
\end{equation}
Defining $\tilde g \equiv - (\dd t + \omega)^2 + h$ as the term in parentheses, the Ricci tensor of \eqref{g rho} is easy to compute:
\begin{equation}
\Ric_g = - 3 \, \dd \rho^2 + \Ric_{\tilde g} - \frac{\varphi_0^2}{L^2} \Big( 2 \sinh^2 \rho + \cosh^2 \rho \Big) \, \tilde g,
\end{equation}
and imposing $\Ric_g = k \, g$ reveals $k = -3/L^2$ as expected.  Plugging in $-(3/L^2) \, g$ on the left, the functions of $\rho$ drop out, and on the 3-metric $\tilde g$ we obtain
\begin{equation} \label{Ric tilde g}
\Ric_{\tilde g} = -2 \frac{\varphi_0^2}{L^2} \, \tilde g.
\end{equation}

Next, we compute the Ricci tensor of $\tilde g \equiv - (\dd t + \omega)^2 + h$, and applying \eqref{d omega}, we obtain
\begin{equation}
\Ric_{\tilde g} = \frac{2}{L^2} \, (\dd t + \omega)^2 + \frac{2}{L^2} \, h + \Ric_h.
\end{equation}
Comparing to \eqref{Ric tilde g}, we must have
\begin{equation}
\varphi_0 = 1, \qquad \text{and} \qquad \Ric_h = - \frac{4}{L^2} \, h.
\end{equation}
But this now uniquely fixes $h$:
\begin{equation}
h = \frac{L^2 \, (\dd x^2 + \dd y^2)}{4 y^2}, \qquad \text{and hence} \qquad \omega = - \frac{L \, \dd x}{2 y}.
\end{equation}
Then the local form of the metric $g$ is completely fixed.  Rescaling $t \to L \, t$, we can write
\begin{equation}
g = L^2 \, \dd \rho^2 + L^2 \, \cosh^2 \rho \bigg[ - \Big( \dd t - \frac{\dd x}{2y} \Big)^2 + \frac{\dd x^2 + \dd y^2}{4y^2} \bigg],
\end{equation}
which is the metric of $AdS_4$ written with $AdS_3$ slices.  Therefore we have shown that the only solutions with timelike Killing vector and at least one supersymmetry are locally AdS.  Applying our boundary conditions, we thus prove that the only asymptotically-AdS ``geon'' with $E = \abs{J}/L$ whose Killing vector is \emph{timelike} in some finite region must in fact be AdS itself.

\subsection{The Null Case}

In the case that the Killing vector $V^\mu \equiv \bar \epsilon \gamma^\mu \epsilon$ is null, the authors of \cite{Caldarelli:2003pb} have obtained the general solution in a form that is easily adaptable to our purpose, so we do not repeat their calculation here.  The result, for vanishing Maxwell field $F$, is given by the metric
\begin{equation} \label{siklos metric}
g = \frac{L^2}{x^2} \Big( \cG(u, x, y) \, \dd u^2 + 2 \dd u \, \dd v + \dd x^2 + \dd y^2 \Big),
\end{equation}
where $V \equiv \partial / \partial v$ and the function $\cG$ must solve
\begin{equation} \label{siklos eqn}
\bigg( \frac{\partial^2}{\partial x^2} + \frac{\partial^2}{\partial y^2} - \frac{2}{x} \frac{\partial}{\partial x} \bigg) \cG(u,x,y) = 0.
\end{equation}
The metric \eqref{siklos metric} has previously been considered by Siklos \cite{Siklos:1985} (a particular example was first found in \cite{Kaigorodov:1963}; for review and analysis see \cite{Podolsky:1997ik}).  One can write down the general solution to \eqref{siklos eqn} by first observing the following identity of operators \cite{Siklos:1985}:
\begin{equation}
\bigg( \frac{\partial^2}{\partial x^2} + \frac{\partial^2}{\partial y^2} - \frac{2}{x} \frac{\partial}{\partial x} \bigg) \bigg( x^2 \frac{\partial}{\partial x} \bigg) \equiv \bigg(x \frac{\partial}{\partial x} - 1 \bigg) \bigg( \frac{\partial^2}{\partial x^2} + \frac{\partial^2}{\partial y^2} \bigg) \Big( x \, \cdot \Big).
\end{equation}
Hence the function $\cG$ can be written
\begin{equation}
\cG(u,x,y) = x^2 \frac{\partial}{\partial x} \bigg( \frac{f(u,z) + \bar f(u, \bar z)}{x} \bigg), \qquad z \equiv x + i y,
\end{equation}
for an arbitrary holomorphic function $f$.

\subsubsection{Applying Boundary Conditions}

Holomorphicity of $f$ is a very strong constraint, and we will show that this is enough to fix $\cG$.  We apply boundary conditions.  The metric \eqref{siklos metric} resembles the Poincar\'e patch of AdS, and so infinity lies at $x \to 0$.  To obtain AdS asymptotics, we must have
\begin{equation}
\cG \to (\text{const}) + \cO(x), \quad \text{as} \quad x \to 0.
\end{equation}
For the function $f$, this corresponds to
\begin{equation} \label{f zero}
f(u, z) \to c_0 + \xi(u) z + \cO(z^2), \quad \text{as} \quad z \to 0,
\end{equation}
where $c_0$ is a constant and $\xi(u)$ is an arbitrary function of $u$.

Likewise, in the limit $x \to \infty$ we demand regularity, which requires
\begin{equation}
\cG \to (\text{const}) + \cO(x^{-1}), \quad \text{as} \quad x \to \infty.
\end{equation}
For the function $f$, this means
\begin{equation} \label{f infty}
f(u, z) \to \tilde \xi(u) z + \tilde c_0 + \cO(z^{-1}), \quad \text{as} \quad z \to \infty.
\end{equation}
But in order for spacetime to be smooth on $0 < x < \infty$, the function $f$ must be holomorphic on $0 < \Re(z) < \infty$.  Thus \eqref{f zero} and \eqref{f infty} must actually be the same Laurent series, and
\begin{equation}
f(u, z) \equiv c_0 + \xi(u) z,
\end{equation}
exactly.  This implies that $\cG = (\text{const})$ everywhere, and thus the metric \eqref{siklos metric} is (locally) uniquely fixed to be AdS.  Applying our global boundary condition, we thus prove that an asymptotically-AdS ``geon'' with $E = \abs{J}/L$ and Killing vector which is \emph{null} in some finite region must in fact be AdS itself.

Thus we prove that in both cases (with either timelike or null Killing field), the unique supersymmetric solution asymptotic to AdS is AdS itself, and there can be no geons with $E = \abs{J}/L$.

\section{Conclusions}

Let us summarise our results.  For any energy and angular momentum in the superradiant regime of Kerr-AdS, there is a countably infinite family of black resonators labeled by an integer $m$.  All of these black resonators have more entropy than Kerr, but are nevertheless still unstable to superradiance.  However, taking the limit $m\rightarrow\infty$ shuts down superradiance and also maximises the entropy.  Therefore, a limiting solution, if it exists, is a natural candidate for the endpoint of the superradiant instability.  But the zero-size limit of these limiting black resonators is expected to be a geon that saturates the minimum energy bound, and is therefore a supersymmetric solution.  Such a geon cannot exist because the only supersymmetric vacuum solution with AdS asymptotics is AdS itself.  

These results support the claim that such a limiting black resonator solution does not exist, since they would not have a zero-size limit.  But, the proof that AdS is the only supersymmetric solution assumes smoothness. There could be a limiting geon that is singular, and it is conceivable for a regular black resonator to cover a singularity with its horizon.  We find this possibility unlikely since the singularity must appear at the centre of the limiting geon in order for a black hole to cover it.  Evidence suggests that the centre of geons becomes smoother (i.e. the curvatures become smaller) rather than more singular in the $m\rightarrow\infty$ limit.  The perturbative results of \cite{Dias:2011ss,Dias:unpublished} found that curvatures in the centre of geons decrease with increasing $m$ for $m=2$, $m=4$, and $m=6$.  Furthermore, the perturbation functions behave near the origin as $r^m$, suggesting that the centre becomes flatter with increasing $m$.  Though, this is not a proof as there may be other factors of $m$ that would affect the curvature.  

Besides the ``limiting" resonators, we know of no other candidate solution for the endpoint of superradiance. If there is indeed no regular solution that is also stable to superradiance, then there are two logical possibilities for the endpoint of the superradiant instability: either a singular solution is reached in finite time, or the system never settles to any solution.  The former would be a violation of cosmic censorship, such as in \cite{Lehner:2010pn}.  In the latter, it is entropically favorable that structure develops on smaller and smaller scales, because black resonators with higher $m$ have higher entropy, and for all $m$ are still unstable to superradiance, as conjectured in \cite{Dias:2011at,Dias:2015rxy}.  At some point, the effects of quantum gravity would become significant, thus violating the spirit, if not the letter, of cosmic censorship.

\newpage

\section*{Acknowledgements}

JES thanks the participants of Strings 2015 for useful feedback, particularly Juan Maldacena. We would like to thank \'O.~J.~Dias for comments on an earlier version of this manuscript.  BEN and BW are supported by ERC grant ERC-2011-StG 279363-HiDGR.




\end{document}